\documentclass[a4paper]{jpconf}
\usepackage{graphicx}
\usepackage{dcolumn}
\usepackage{bm}
\usepackage{rotating}

\begin{document}

\title{Single-particle isomeric states in $^{121}$Pd and $^{117}$Ru}

\author{S~Lalkovski$^{1,2,\dagger}$, A~M~Bruce$^2$, A~M~Denis~Bacelar$^2$, 
M~G\'orska$^3$, S~Pietri$^{3,4}$, Zs~Podoly\'ak$^4$, P~Bednarczyk$^3$, L~Caceres$^3$, 
E~Casarejos$^5$, I~J~Cullen$^4$, P~Doornenbal$^{3,6}$, G~F~Farrelly$^4$, A~B~Garnsworthy$^4$, 
H~Geissel$^3$, W~Gelletly$^4$, J~Gerl$^3$,  J~Gr\c{e}bosz$^{3,7}$, C~Hinke$^{8}$, G~Ilie$^{9,10}$, 
G~Jaworski$^{11,12}$, S~Kisyov$^1$, I~Kojouharov$^3$, N~Kurz$^3$, S~Myalski$^7$, M~Palacz$^{12}$, 
W~Prokopowicz$^3$, P~H~Regan$^4$, H~Schaffner$^3$, S~Steer$^4$, S~Tashenov$^3$,  P~M~Walker$^{4,13}$, 
H~J~Wollersheim$^3$ and M~Zhekova$^1$}

\address{
$^1$Faculty of Physics, University of Sofia "St. Kliment Ohridski'', Sofia 1164, Bulgaria\\
$^2$School of Computing, Engineering and Mathematics, University of Brighton, Brighton BN2 4JG, UK\\
$^3$Gesellschaft f\"ur Schwerionenforschung mbH, Planckstr 1, D-64291 Darmstadt, Germany\\
$^4$Department of Physics, University of Surrey, Guildford GU2 7XH, UK\\
$^5$Facultad de F\'isica, Universidad de Santiago de Compostela, Santiago
de Compostela 15782, Spain\\
$^6$Institut f\"ur Kernphysik, Universit\"at zu K\"oln, Z\"ulpicher Stra$\ss$e 77,
D-50937 K\"oln, Germany\\  
$^7$Niewodnicza\'nski Institute of Nuclear Physics, Polish Academy of
Science, ul. Radzikowskiego 152, Krakow 31-342, Poland\\
$^8$Physik-Department E12, Technische Universit\"at M\"unchen, D-85748
Garching, Germany\\
$^9$Wright Nuclear Structure Laboratory, Yale University, New Haven, Connecticut
06520, USA\\
$^{10}$National Institute for Physics and Nuclear Engineering, P.O. Box MG-6,
Bucharest, Romania\\
$^{11}$Heavy Ion Laboratory, Warsaw University, ul. Pasteura 5A, 02-093 Warszawa, Poland\\
$^{12}$Faculty of Physics, Warsaw University of Technology, Koszykowa 75,
00-662 Warszawa, Poland\\
$^{13}$CERN, CH1211 Geneva 23, Switzerland\\
}
\ead{$^\dagger$stl@phys.uni-sofia.bg}

\date{\today}

\begin{abstract}
Neutron-rich nuclei were populated in a relativistic fission of $^{238}$U. 
Gamma-rays with energies of 135 keV and 184 keV were associated with two isomeric states 
in $^{121}$Pd and $^{117}$Ru. Half-lives of 0.63(5) $\mu$s  and 2.0(3) $\mu$s were deduced 
and the isomeric states were interpreted in terms of deformed single-particle states.
\end{abstract}

\section{Introduction}
Nuclear shell structure is one of the milestones in nuclear physics.
It is related to enhanced robustness of nuclei with respect to excitations, 
when a particular number of nucleons is present \cite{MGM48}. 
A key element of the nuclear shell model is the spin-orbit force, 
which lifts the $j$-degeneracy and pushes down the orbits with higher-$j$ 
towards orbits of opposite parity. One of the major successes of the 
model is the ability to reproduce the nuclear magic numbers and to 
explain the islands of isomerism emerging on the Segr\'e chart when approaching
the magic medium-mass and heavy nuclei. 

In addition to the spin-orbit force, it is the nuclear quadrupole deformation, 
which splits the high-$j$ states into a number of orbits with a different third 
projection of the total angular momentum \cite{Ni55}, causing new sub-shell closures 
to emerge \cite{XWW02} and high-$K$ isomeric states to appear in the deformed regions 
on the Segr\'e chart \cite{WD99}.

The present article addresses the structure of $^{121}_{\ 46}$Pd$_{75}$ and 
$^{117}_{\ 44}$Ru$_{73}$, which are the most neutron-rich nuclei in the palladium 
and ruthenium isotopic chains studied by means of $\gamma$-ray spectroscopy. They
are placed in a region of the Segr\'e chart where variety of shapes are expected, 
involving large degree of triaxiality and $\gamma$-softness \cite{SMN97, NSO10}. Also, 
because of the vicinity of the $N=82$ magic number, a deformed-to-spherical shape 
transition \cite{SMN97, NSO10} is expected to take place. Due to the nature of the 
shell structure, isomeric states were observed in the two nuclei, which can be related  
to specific single-particle configurations, and hence can shed light on the nuclear shapes in
the mass region.

\section{Experimental set up and data analysis}

Neutron-rich palladium and ruthenium isotopes were produced in a relativistic fission 
reaction. A $^{238}$U beam was accelerated up to 750 MeV/n by the GSI SIS accelerator and 
impinged on 1g/cm$^2$ $^9$Be target. The fission products were separated by the GSI 
Fragment Separator (FRS) and implanted in a passive stopper. Isomeric delayed 
transitions were detected by the RISING detector array \cite{Pi07}, comprising 105 HPGe 
detectors. The data were processed by XIA DGF modules providing signals with an energy 
resolution of 3 keV at 1.3 MeV and a time resolution of 25 ns. The data acquisition was 
triggered by a particle detector, placed at the final focal plane of the FRS.
The data were stored in event-by-event mode. Experimental details were previously
published elsewhere \cite{Br10}.

Fig.~\ref{ene121} presents the $\gamma$-ray energy spectrum in singles, obtained in coincidence 
with the $^{121}$Pd nuclei. A $\gamma$-transition with an energy of 135 keV is observed.
The inset of the figure presents the time distribution, gated on the 135-keV $\gamma$-ray. The 
half-life, obtained with a single exponential curve, is 0.63(5) $\mu$s. 

\begin{figure}[t]
\begin{minipage}{18pc}
\rotatebox{-90}{\scalebox{0.28}[0.28]{\includegraphics{Ene_time_Pd121.epsi}}} 
\caption[]{\label{ene121} Isomeric transition in $^{121}$Pd and time spectrum (inset) gated on it.}
\end{minipage}\hspace{2pc}
\begin{minipage}{18pc}
\rotatebox{-90}{\scalebox{0.28}[0.28]{\includegraphics{Ene_time_Ru117.epsi}}} 
\caption[]{\label{117ru} Isomeric transition in $^{117}$Ru and time spectrum (inset) gated on it.}
\end{minipage}
\end{figure}
Fig.~\ref{117ru} shows the energy spectrum, gated on $^{117}$Ru ions, where a single transition 
with an energy of 184 keV was observed. The inset of the figure presents
the time distribution of the 184-keV transition and a half-life of 2.0(3) $\mu$s was 
obtained from the slope of the curve. The two $\gamma$-rays in $^{121}$Pd and $^{117}$Ru were  
previously observed \cite{TMW07}, but no half-life information was published.

\section{Discussion}
\begin{figure}[t]
\begin{center}
\rotatebox{90}{\scalebox{0.53}[0.53]{\includegraphics{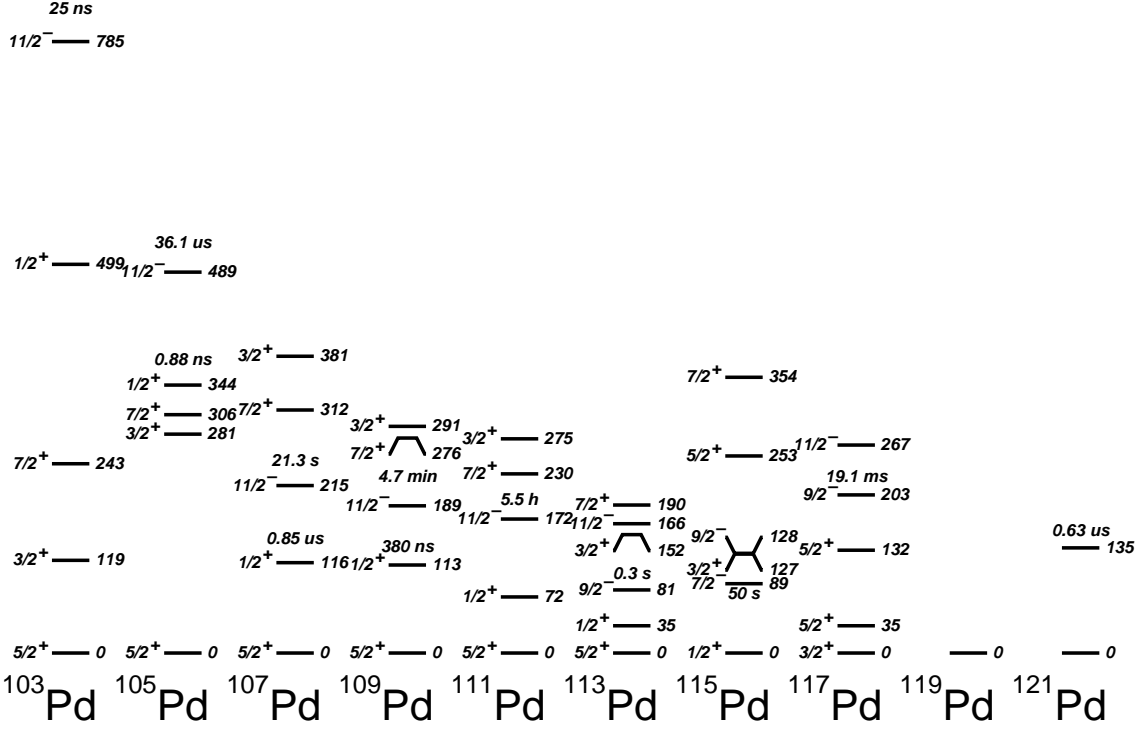}}} 
\caption[]{\label{oddPd}Systematics of the low-lying excited states in the neutron-rich palladium isotopes:  $^{103-113}$Pd \cite{nndc}, $^{115}$Pd \cite{Ku10}, $^{117}$Pd \cite{Ur04} and $^{121}$Pd (present work).}
\end{center}
\end{figure}

\begin{figure}[t]
\begin{minipage}{18pc}
\rotatebox{-90}{\scalebox{0.4}[0.4]{\includegraphics{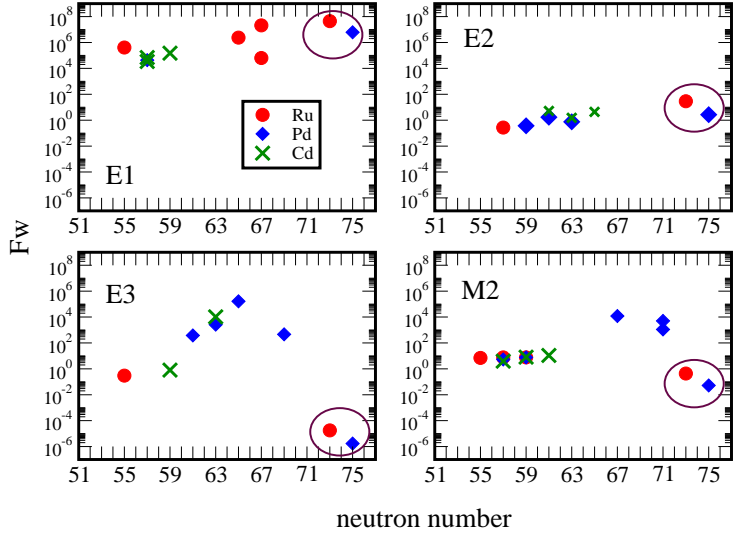}}} 
\end{minipage}\hspace{8pc}
\begin{minipage}{10pc}
\caption[]{\label{Fw_sys} Hindrance-factor systematics for the low-lying isomeric states 
in the neutron-rich odd-$A$ ruthenium, palladium and cadmium nuclei. The encircled regions 
present the hindrance factors for the 135-keV and 184-keV transitions studied here.}
\end{minipage}
\end{figure}

Fig.~\ref{oddPd} presents the systematics of the low-lying excited states 
in the neutron-rich odd-$A$ palladium isotopes. The ground state in $^{103}$Pd 
to $^{113}$Pd is a $J^\pi(g.s.)=5/2^+$ state \cite{nndc},
while the $J^\pi$ value for the ground state in $^{115}$Pd and $^{117}$Pd are subject of 
discussion in the literature \cite{nndc, Ku10, Ur04}. Here, we adopt the $J^\pi$ 
assignments for $^{115}$Pd and $^{117}$Pd as given in \cite{Ku10} and \cite{Ur04}, respectively.
The extrapolation of the systematics towards the extremely 
neutron-rich odd-$A$ palladium nuclei suggests that $J^\pi=1/2^+$ to $7/2^+$ and $J^\pi=7/2^-$ to 
$11/2^-$ states will be present close to the $^{121}$Pd ground state, but not all candidates give 
rise to a sub-microsecond isomer. The low-lying 
isomeric states, observed in the odd-$A$ palladium isotopes are related to the decay of the 
negative-parity states as well as to the decay of the $J^\pi=1/2^+$ excited state via stretched $E2$ 
low-energy transitions. Hindrance factors (HF), defined as $F_W=T_{1/2,\gamma}/T_{1/2,W.e.}$ where 
$T_{1/2,\gamma}$ is the partial half-life and $T_{1/2, W.e.}$ is the single-particle estimate, were 
calculated for the isomeric $E1$, $E2$, $E3$ and $M2$ transitions and  plotted on 
Figure~\ref{Fw_sys} as a function of the neutron number.

\begin{figure}[ht]
\begin{center}
\rotatebox{90}{\scalebox{0.53}[0.53]{\includegraphics{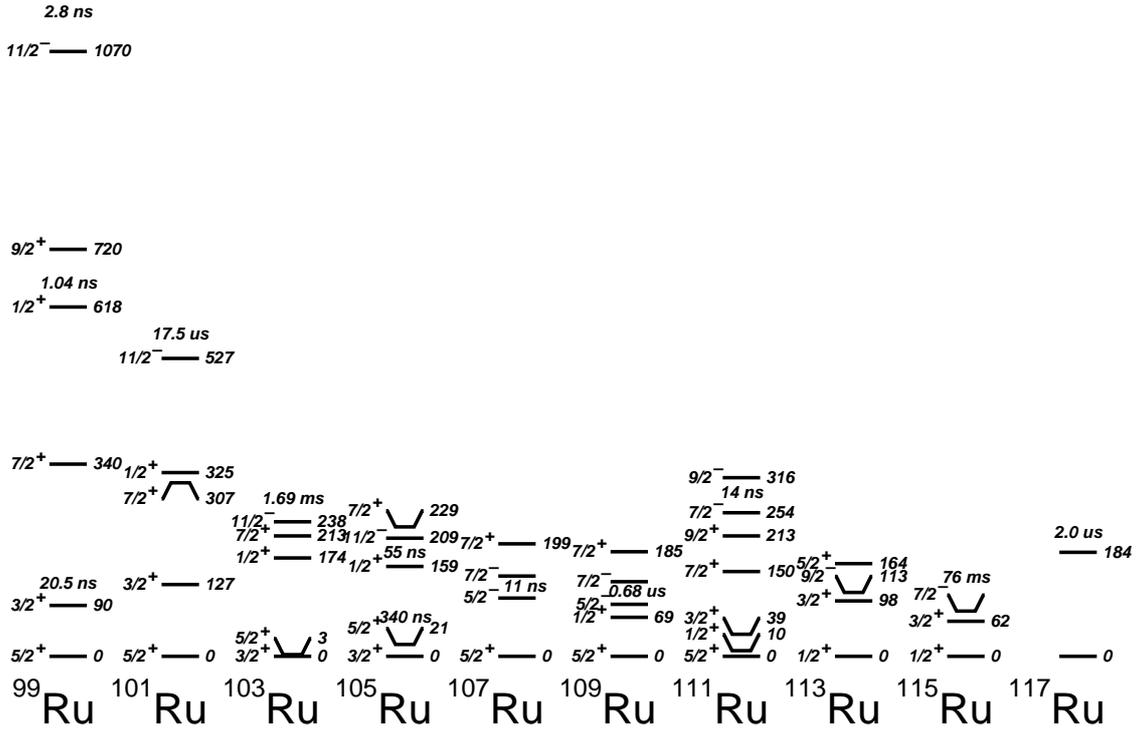}}} 
\caption[]{\label{oddRu}Systematics of the low-lying excited states in the neutron-rich ruthenium isotopes. Data from \cite{nndc} for $^{99-115}$Ru and from the present work for $^{117}$Ru. }
\end{center}
\end{figure}

The neutron-rich odd-$A$ palladium ($Z=46$) isotopes show structure close to that of the 
odd-$A$ ruthenium ($Z=44$) nuclei (Figure~\ref{oddRu}). The HF, calculated for 
the low-lying isomeric transitions in the neutron-rich odd-$A$ ruthenium nuclei are also plotted on 
Fig.~\ref{Fw_sys}. To increase the statistics on the systematics in Figure~\ref{Fw_sys}, 
HF were calculated for a number of low-lying isomeric transitions in the odd-$A$ Cd  
nuclei \cite{nndc, Ki11}. Inside the encircled regions on Figure~\ref{Fw_sys}
HF for the 184-keV and 135-keV isomeric transitions in $^{117}$Ru and $^{121}$Pd are presented.

On Figure~\ref{Fw_sys}, the $E1$ HF are in the $10^4 < F_W < 10^8$ range. If the 135-keV and 
the 184-keV isomeric transitions, observed in $^{121}$Pd and $^{117}$Ru nuclei, are of $E1$ nature 
the respective HF will follow the systematics presented in Figure~\ref{Fw_sys}. The $E2$ 
HF shows a steady behaviour with the neutron number having $F_w\approx 1$. The $^{121}$Pd and 
$^{117}$Ru data points would follow the systematics in Figure~\ref{Fw_sys} if the two 
transitions are of $E2$ nature. In contrast to the $E2$ HF, the $E3$ $F_W$ data points 
show an abrupt change varying six orders of magnitude. If the 135-keV and 184-keV transitions
are of $E3$ nature, then they will be six orders of magnitude more enhanced than the other $E3$ 
transitions. Hence, the$E3$ nature for the two isomeric transitions in $^{117}$Ru and $^{121}$Pd 
can be ruled out. The $M2$ transitions in the region 
have $F_w\geq 1$. If 135-keV and 184-keV transitions are of $M2$ nature, than they would 
be enhanced and would not follow the systematics of the $M2$ transitions. Therefore, $M2$ 
nature is unlikely.
  
The single-particle levels, calculated with a Woods-Saxon potential for the nuclei 
in the $A\approx 110$ mass region region \cite{XWW02}, show that the $^{121}$Pd Fermi 
surface involves $7/2^+[404]$, $1/2^+[400]$, $3/2^+[402]$, and $9/2^-[514]$ Nilsson orbits, 
where $\beta_2 =0.19$ was calculated from the energy of the first excited state in $^{120}$Pd. 
Given the HF systematics (Fig.~\ref{Fw_sys}), $9/2^-\rightarrow 7/2^+$ $E1$ and 
$3/2^+\rightarrow 7/2^+$ $E2$ transitions would cause isomeric states in the sub-microsecond 
time range. However, $J^\pi=9/2^-$ assignment for the isomeric state can be ruled out because 
of the direct $\beta$-feeding observed to the $9/2^+$ state in $^{121}$Ag \cite{St09} suggesting 
the $J^\pi=9/2^-$ isomeric state in $^{121}$Pd has a half-life longer than the half-life of the 
observed in the present study nanosecond isomer. Therefore, we tentatively interpret the 135-keV 
transition as a prolate $3/2^+[402] \rightarrow 7/2^+[404]$  $E2$ transition.

The 73$^{th}$ neutron of  $^{117}$Ru, which has $\beta_2=$0.25 estimated from the first excited state
in $^{116}$Ru, is placed on $1/2^+[400]$ Nilsson orbit \cite{XWW02}. At this deformation, the 
$1/2^-[541]$ down-sloping orbit is also present close to the ground state, therefore we tentatively 
interpret the isomeric transition as a $1/2^-\rightarrow 1/2^+$ $E1$ transition. It should be noted, 
however, that at this deformation the $3/2^+[402]$, $7/2^+[404]$ and $9/2^-[514]$ Nilsson orbits are 
also present close to Fermi level, which can give rise to a situation similar to the $^{121}$Pd 
scenario. The analysis of $^{121}$Pd and $^{117}$Ru is further complicated by the fact that the 
two nuclei are in a region where triaxiality, shape transitions and shape co-existence \cite{NSO10}
are expected to occur, which may result in unexpected deviations from the hindrance factors 
systematics presented here. Therefore, in order to draw firm conclusions on the structure of 
these neutron-rich nuclei more experimental information and deeper theoretical analysis are needed. 

\section{Conclusions}

Neutron-rich nuclei were populated in a relativistic fission of $^{238}$U. 
Gamma-rays with energies of 135 keV and 184 keV were associated with  
isomeric states in $^{121}$Pd and $^{117}$Ru respectively. Half-lives of 0.63(5)$\mu$s 
and 2.0(3)$\mu$s were deduced and the isomeric states are interpreted 
in terms of prolate deformed single-particle states.

\ack
This work is partially supported by the UK STFC, AWE plc., Royal Society, and by 
the Bulgarian National Science Fund under contract No: DMU02/1. S.L. acknowledges 
constructive discussions with Dr. F.G. Kondev and Prof. F.R. Xu.

\end{document}